\documentclass[twocolumn]{aa}
\usepackage{graphicx}
\usepackage{txfonts}
\begin{document}

   \title{Near infrared imaging of \object{NGC\,2316}\thanks{Based on observations carried out at ESO, La Silla, Chile.}}

   \author{P. S. Teixeira
          \inst{1}
          \and
          S. R. Fernandes\inst{1}
	  \and
	  J. F. Alves\inst{2}
	  \and
	  J. C. Correia\inst{1}
	  \and
	  F. D. Santos\inst{1}
	  \and
	  E. A.  Lada\inst{3}
	  \and
	  C. J.  Lada\inst{4} 
          }

   \offprints{P. S. Teixeira}

   \institute{Depart. de F\'{\i}sica, Faculdade de Ci\^encias da 
Universidade de Lisboa, 
Ed. C8, Campo Grande, 1749-016 Lisboa, Portugal\\
              \email{psteixeira@fc.ul.pt, srfernandes@fc.ul.pt, 
jcc@fc.ul.pt, fdsantos@fc.ul.pt}
         \and
              European Southern Observatory, ESO, Germany\\
             \email{jalves@eso.org}
	 \and
	     Department of Astronomy, University of Florida, U.S.A.\\
	     \email{lada@astro.ufl.edu}
	 \and
	     Harvard-Smithsonian Center for Astrophysics, CfA, U.S.A.\\
	     \email{clada@cfa.harvard.edu}
             }

   \date{}

   \abstract{In the present paper we present $JHK$ photometric results of the young embedded cluster NGC\,2316. We construct the cluster radial profile from which we determine a radius of 0.63\,pc. We find 189\,$\pm$\,29 cluster members in an extinction limited sub-sample of the survey, 22\,$\pm$\,19 of which are possibly substellar. An average extinction of 4.5 visual magnitudes is derived using $(H - K)$ colours of control fields. This extinction is due to the presence of residual parental molecular cloud. NGC\,2316 presents 16\% source fraction of excess emission which is consistent with other results from clusters with an age of 2 -- 3\,Myr. This age is consistent with the distribution of sources in the colour-magnitude diagram when compared to theoretical isochrones, and the overall shape of the cluster KLF. The substellar population of the cluster is similar or smaller than that observed for other embedded clusters and the stellar objects dominate the cluster membership.

   }

   \maketitle
%

\section{Introduction}
Embedded clusters may be the fundamental units of star for\-ma\-tion (Lada \& Lada, 2003). They contain statistically si\-gni\-fi\-cant samples of young stars of similar age and composition spanning a wide range of mass, providing excellent la\-bo\-ra\-to\-ries for investigating a number of important issues in star formation such as the form and universality of the initial mass function (IMF) and the frequency and lifetimes of protoplanetary disks. Surveys for near infrared (NIR) excess in embedded clusters can provide knowledge of the disk frequency and evolution because a protoplanetary disk is easier to detect than a planetary system with a similar mass of solid material (Beckwith \& Sargent, 1996). In addition, embedded clusters provide the smallest spatial scale for investigating the nature of the IMF, its form and universality; their mass func\-tion is in fact an initial mass function since young clusters have not lost significant numbers of members to either dynamical or stellar evolution.\\NGC\,2316 is a young partially embedded cluster with coordinates $(\alpha,\delta)(J2000)=(6^\mathrm{h}59^\mathrm{m}40^\mathrm{s},-7\degr46\arcmin36\arcsec)$ at a distance of 1.1\,kpc according to Felli et al. (1992) and Hodapp (1994). There is an HII region centered on a B3 ZAMS star (Noguchi et al., 1993; Fukui et al., 1993),  associated with the \object{IRAS 06572-0742} source. The UV radiation is apparently creating a dense photodissociated region (Ryder et al., 1998), showing some physical signatures of ongoing star formation, such as $H_2O$ masers (Felli et al., 1992) and $CO$ outflow with spectral index characteristic of optically thin free-free emission (Beltr\'an et al., 2001).
NGC\,2316 was first uncovered by Parsamian (1965), followed by other studies (e.g. Hodapp, 1994), however certain parameters are still not well established. To better characterize this cluster we obtained deep multi wavelength NIR images of the region to better constrain the cluster's age and spatial distribution, as well as provide estimates of membership number, fraction of NIR excess sources and brown dwarf po\-pu\-la\-tion.

\section{Observations}
\begin{figure}[!h]
  \centering
  \caption{(See attached file 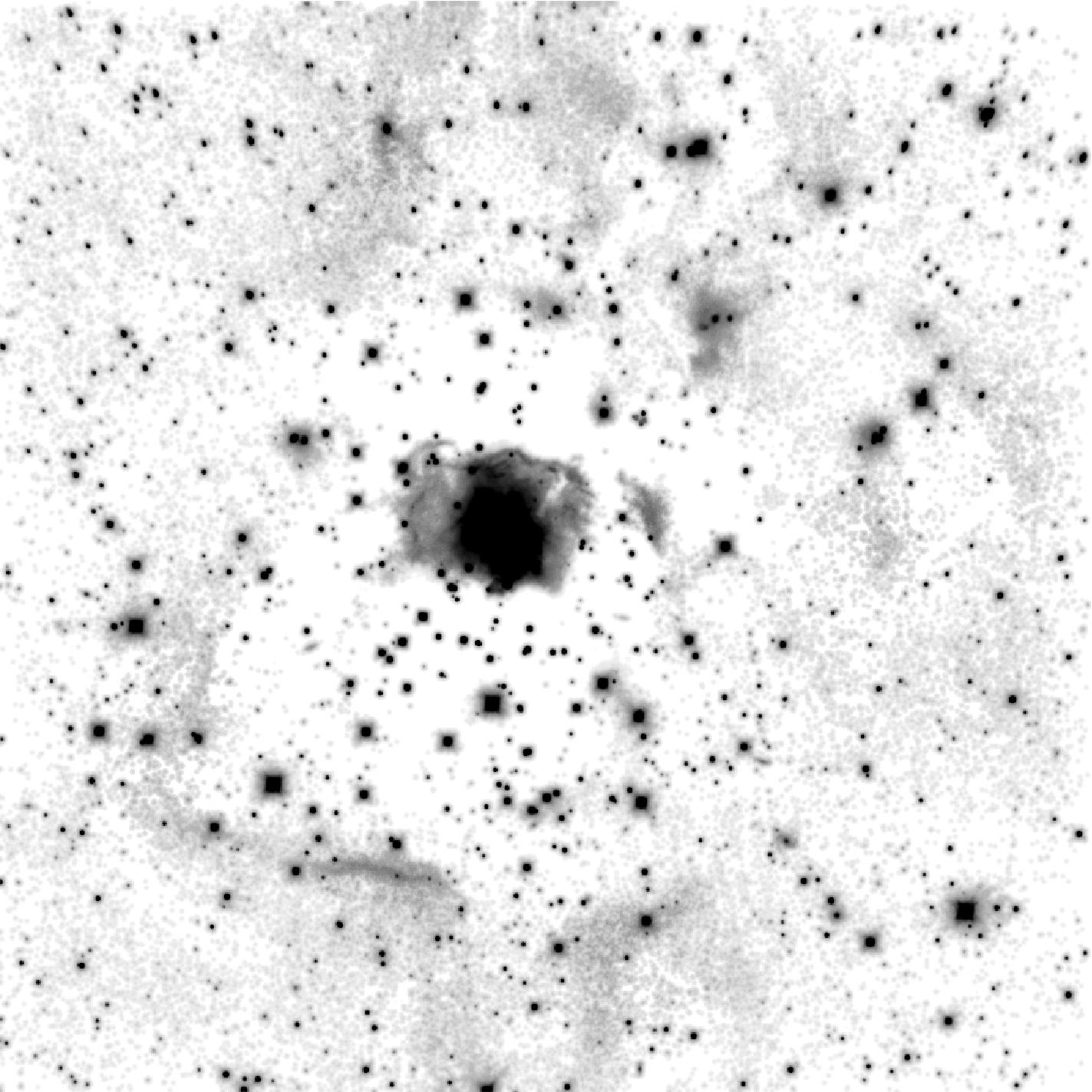) NGC\,2316 $JHKs$ combined grey scale image from the NTT data.The scale is stretched to enhance the low level emission due to residual cloud material. The large majority of sources in this image are not visible at optical wavelengths. The field size is 5\arcmin$\times$5\arcmin.}
  \label{fotosL1654}
\end{figure}
 The observations were made  on the 8th and 9th of March 1999 using the 
SofI infrared camera ($J\, (1.25\mu$m), $H\, (1.65\mu$m), and $Ks\, 
(2.162\mu$m) bands) on the 3.5\,m NTT telescope in La Silla, Chile. We took 45 
images per band, with a total integration time of 900\,s.
Two control fields were taken also in $JHKs$ bands, one located 20\arcmin\, N of the NGC\,2316's 
center, the other located 13\arcmin\,S, with a total integration time of 90\,s.
The field of view was aproximately 5\arcmin\, $\times$\, 5\arcmin, with 
a pixel scale of\, 0.29\arcsec per pixel. The average FWHM of point sources in the final i\-ma\-ge is about 0.54\arcsec.
In Fig. 1 we can observe that NGC\,2316 presents 
a residual cloud material, nicely visible as a bubble centered on the cluster's core. This faint nebulae, surrounding the B star in the center of the cluster, is indicative 
of the early age of this cluster, less than 5\,Myr when compared to NGC\,2362 (\cite{moitinho}) which shows no nebulosity.
\section{Data analysis}
The data reduction for NGC\,2316 cluster is done using the IRAF\footnote{\emph{Image 
Reduction and Analysis Facility} (IRAF) is distributed by NOAO, which is operated by AURA, Inc., under contract to the NSF.} DIMSUM (Deep Infrared Mosaicing Software) 
pa\-ckage, to reconstruct dithered exposures taken for $J$, $H$ and $Ks$ 
filters.
Astrometry is conducted with ESO's (European Southern Observatory) 
SKYCAT tool and the IRAF routines IMTRANSPOSE, 
CCMAP and CCSETWCS.
Infrared sources are identified u\-sing SExtractor (\cite{Bertin}) and 
checked through visual inspection.  A FWHM of 0.5\arcsec is used, along 
with a gaussian filter of 3 sigma, a detection threshold of 3 sigma and a 
deblending threshold parameter of 64 with a mi\-ni\-mum contrast of 0.0005 
counts. We use the SExtractor output to remove approximately 80 galaxies from the cluster and control fields $K$--band data.
Photometry is done with the routine PHOT of the IRAF's PHOTCAL 
package, u\-sing an aperture radius of 5 pixels (0.73\arcsec) for the cluster 
and 6 pixels (0.87\arcsec) for the two control fields with appropriate 
aperture corrections per band.
Finally, for the photometric calibration we reduced 9 exposures of the 
standard star 9118 from the Persson catalog (\cite{Persson}), 
determining the zero point by using the filter transformation equations in the 
SofI's camera web page\footnote{http://www.ls.eso.org/lasilla/Telescopes/NEWNTT/sofi/.}. We converted our $Ks$ photometric results to $K$ fotometry using these equations to better compare with other similar studies on young clusters. Comparing our photometry with 2MASS photometry\footnote{This publication makes use of data products from the Two Micron All Sky Survey, which is a joint project of the University of Massachusetts and the Infrared Processing and Analysis Center/California Institute of Tecnology, funded by the National Aeronautics and Space Administration and the National Science Foundation.}, we find an average difference in $J$, $H$ and $K$ magnitudes of 0.04, 0.02 and 0.04, respectively and an average difference in $(H-K)$ colours of -0.01 and in $(J-H)$ of 0.03, so no further calibration is needed.
\section{Results}
In the present $JHK$ photometry survey of NGC\,2316, we detected 1067 
sources at 
$J$--band, 1229 sources at\, $H$--band , and 1251 sources at 
$K$--band. 
The completeness limits of the observations are 18.0, 18.5 and 20.0 for the $K$, $H$, and $J$--bands, respectively, in the cluster field, and 17.5, 17.5, 18.0 for an a\-ve\-raged control field. These are estimated from the luminosity functions, where we consider  our sample complete up to the bin imediately before the maximum. For the data analysis the control field completeness limits were used and no extrapolation was made to account for sensitivity diferences between the two fields observed.
\subsection{Radial profile}
\begin{figure}
  \caption{(See attached file 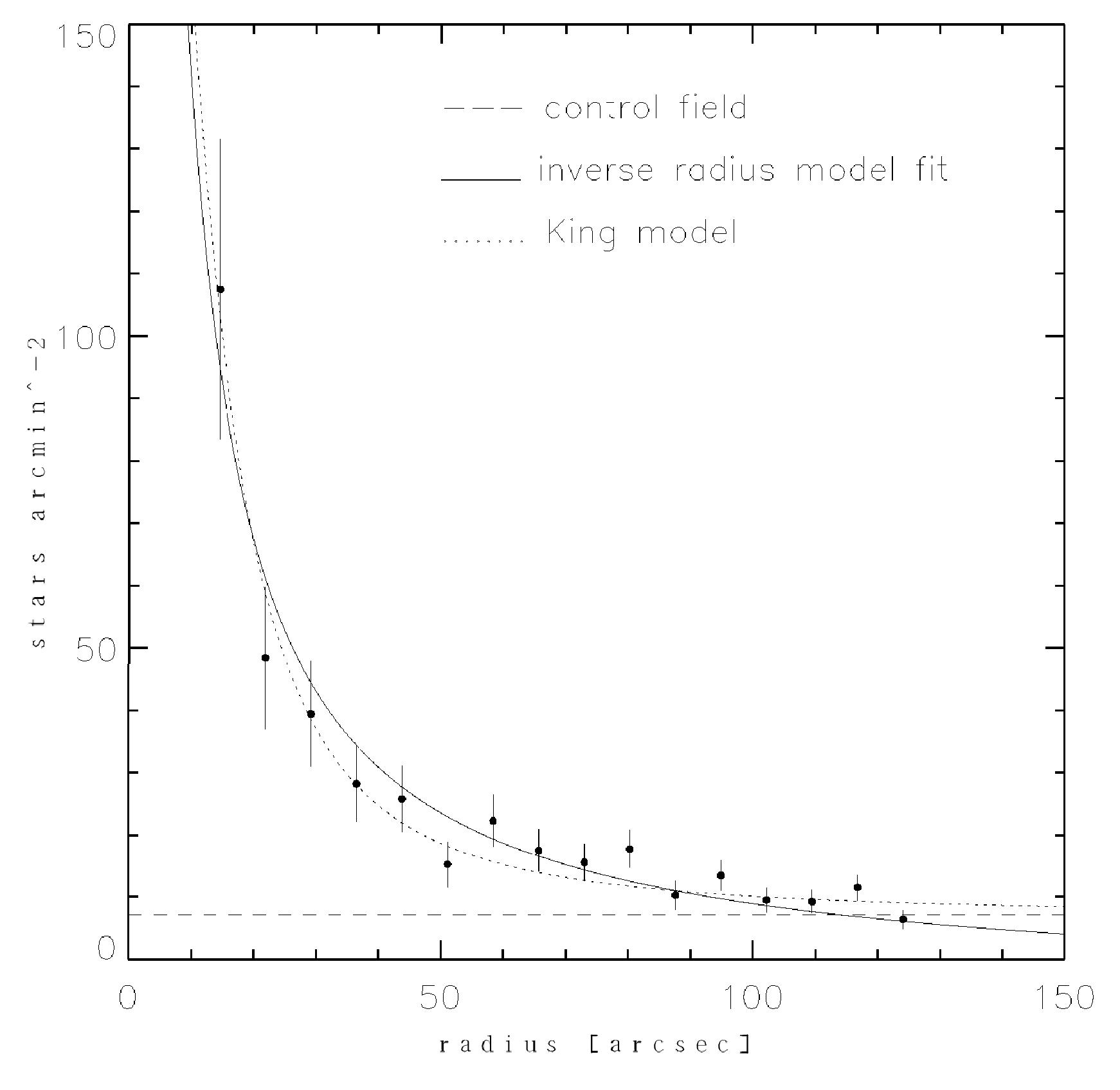) $K$-band radial profile of NGC\,2316.
The error bars represent the $\sqrt{N}$ statistical error in each bin.
The dashed line corresponds to the background density determined from 
averaging the reddened ($\langle$A$_\mathrm{V}\rangle$=4.5 magnitudes) control field. Two fits are plotted, 
one using the King (1962) model, the other an inverse 
radius profile.
From this we estimate the cluster radius to be 120\arcsec, or 
0.63\,pc.}
  \label{RadialProfile}
\end{figure}
Figure 2 presents the $K$ radial density profile, where only the sources brighter than the completeness limit were included. 
The profile is  centered on the peak of the stellar density distribution. NGC\,2316 presents a regular distribution
falling steeply and merging with the extincted background at 
120\arcsec (0.63\,pc). The stellar surface density distribution exhibits more of a centrally condensed and relaxed structure than a hierarchical type structure (e.g., Lada \& Lada, 2003). Within the 120\arcsec\,radius, the number of cluster members is 158$\pm$20. Assuming a spherical geo\-me\-try and an average mass per star of 0.3\,M$_{\odot}$, we determine  the stellar volume density to be 45\,$\mathrm{M}_{\odot}\,\mathrm{pc}^{-3}$.
Two models are fitted to our radial profile as shown in the figure. 
The King model (\cite{king}) is given by the expression $n(r)=f_0/[1+(r/r_\mathrm{c})^2]$, where $f_0$ is the core concentration at zero radius and $r_\mathrm{c}$ is the core radius. This function agrees sufficiently with our data ($\chi^2=0.81$) to provide an estimate of the core radius and density. The parameters are determined to be 300 stars arcmin$^{-2}$ (72 M$_\odot$\,pc$^{-3}$) and 10\arcsec\ (0.05\,pc) respectively for $f_0$ and $r_\mathrm{c}$. An inverse radius model ($n(r)=-5.77 + 1466.61/r$, $\chi^2=0.94$) also des\-cri\-bes fairly well the density distribution. We also note some oscilations in the density distribution that can be of statistical origin.
\subsection{Colour-colour diagram}
\begin{figure}
  \caption{(See attached file 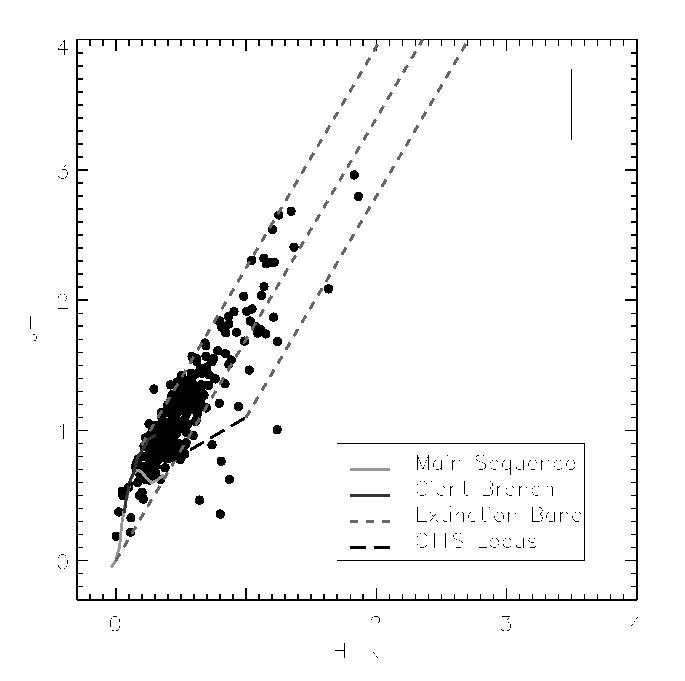)$(H-K)$ vs. $(J-H)$ colour-colour diagram for NGC\,2316 sources. The giant branch and main sequence from Bessell \& Brett (1988) 
are shown, along with the Classical T-Tauri locus from Meyer et al. (1977). The 
reddening vector was determined using the va\-lues of extinction
from Rieke \& Lebofsky (1985). Photometric errors are illustrated at the upper right.}
  \label{CCdiagram}
\end{figure}
The infrared $JHK$ colour-colour diagram of NGC\,2316 is plotted on Fig. 3, considering only sources brighter than the completeness limits for each band and located inside the cluster radius, as determined in the previous section. There is a considerable spread of stars along the reddening band, which corroborates with the previous statement that the cluster is partially embedded. $(J-H)$ and $(H-K)$ colours are also calculated for the two control fields for comparison. The latter fields show almost no reddening. Subtracting this contamination we find 16$\% \pm$ 3$\%$ of cluster members with NIR excess emission characteristic of young stars with circumstellar disks (Lada \& Adams, 1992).
\subsection{Colour-magnitude diagram}
\begin{figure}
  \caption{(See attached file 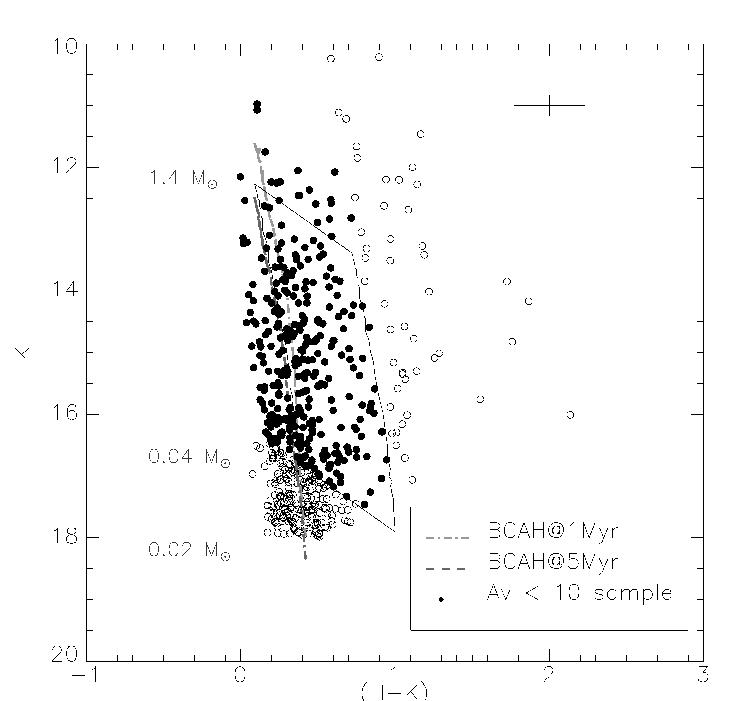)$K$ vs. $(H-K)$ colour-magnitude diagram for the entire survey.  The data are compared with the 1 and 5 Myr isochrones of Baraffe et al. (1998) at a distance of  1.1 Kpc. A sample is selected for sources with A$_\mathrm{V} <$ 10; the reddening slope is determined from Rieke \& Lebofsky (1985).}
  \label{CMdiagram}
\end{figure}
The infrared colour-magnitude diagram for NGC\,2316 is presented in 
Fig. 4 for sources with magnitudes above the completeness limits. We compare source
 locations with the theo\-re\-ti\-cal isochrones of 1 and 5 Myrs from the 
Baraffe et al. (1998) non-grey evolutionary models at a distance of 1.1 Kpc (\cite{Felli}). 
 There is some spread of sources to the left of the isochrones due to foreground contamination.
From the diagram, we observe a much wider spread of sources to the right of the isochrones, a result of extinction produced by the associated molecular cloud. In order to derive the cluster KLF, we construct a statistically more complete mass and extinction sample by including only objects with A$_\mathrm{V} <$ 10 magnitudes and mass $>$ 0.04\,M$_\odot$, for which we are essentially complete. This diagram is not a good discriminant for the cluster age so we assume an intermediate age of 
3 Myrs for the construction of our mass limited and extinction limited sample.Within the errors this sample is also valid for
cluster ages
of 1-5Myr.
\subsection{Luminosity function}
The absolute $K$ luminosity function of the cluster 
(KLF), adjusted for extinction, is plotted in Fig. 5, where the hydrogen burning limit is indicated by a vertical line. Here we only use sources from the sample determined from the colour-magnitude diagram.
 This KLF is determined by subtracting a reddened control field from the original reddened on field. The off fields are averaged into a final control field and scaled so that the control field and the on field have both the same area. The extinction is determined from the average colour excess of the NGC\,2316 field as compared to the control field ($\Delta(H-K)=\langle(H-K)_\mathrm{cluster}\rangle - \langle(H-K)_\mathrm{control field}\rangle$) using sources with no NIR excess emission. This average colour excess is used to determine the extinction following the Rieke \& Lebofsky (1985) law and is found to be $\langle A_\mathrm{V}\rangle =4.5$ magnitudes with a standard deviation of approximately 5 magnitudes. We compare this luminosity function with those of Trapezium (\cite{Muench02}) and \object{IC\,348} (\cite{lisa}), where we find a striking similarity between the shapes of NGC\,2316 and IC\,348 KLFs. We estimate a membership of 189\,$\pm$\,29 and a number of objects beyond the HBL of 22\,$\pm$\,19. This is a very crude estimate because although we are sensitive to masses as low as 20\,M$_\mathrm{Jup}$, the background contamination dominates the statistical errors for the substellar KLF bins at NGC2316's distance.
\begin{figure}
  \caption{(See attached file 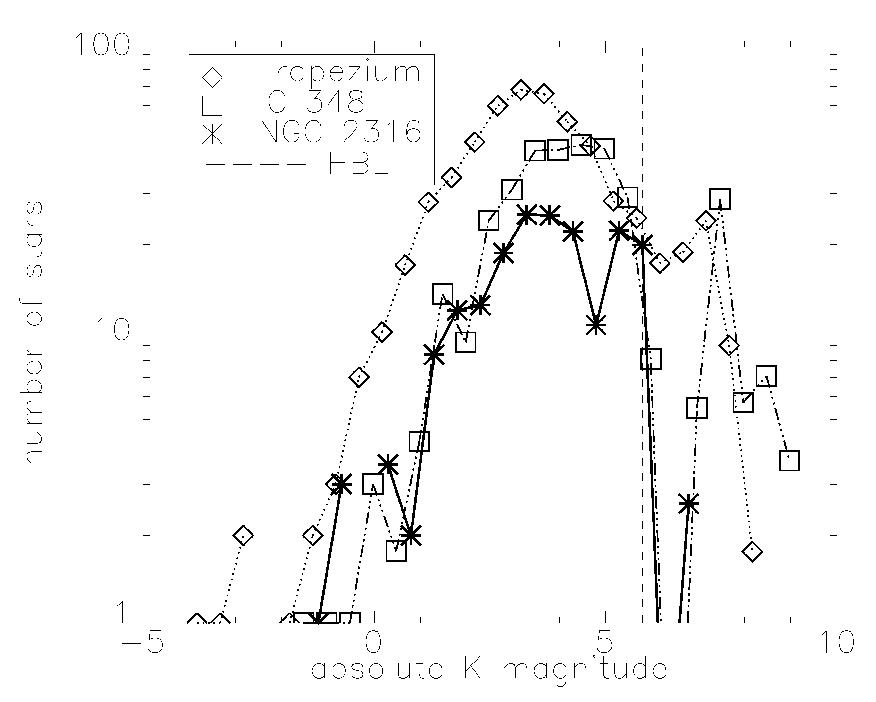)Dereddened KLF of NGC2316, compared to the KLFs of the \object{Trapezium} and IC\,348 clusters. The vertical dotted line indicates the hydrogen burning limit (HBL), (\cite{Baraffe}) for NGC\,2316 assuming a distance of 1.1\,kpc and an age of 3\,Myr.}
  \label{KLF}
\end{figure}
\section{Discussion and summary}
Circumstellar disks emit at NIR wavelengths. Studying the NIR excess 
emission of young cluster sources provides thus information on the 
frequency of circumstellar disks. Our estimate of the NIR excess 
emission for NGC\,2316 according to the 
colour-colour diagram is 16\% $\pm$3\%.\\
We conclude that NGC\,2316's members 
span a wide range of masses, from B--stars down to the substellar regime (40\,M$_\mathrm{Jup}$), and is 
significantly embedded ($\langle A_\mathrm{V}\rangle\,=\,4.5$) in its parental molecular cloud. 
We estimate a size of 0.63\,pc for NGC\,2316 from the radial profile assuming a distance of 1.1\,kpc. Using an extinction limited sub-sample of the survey, we derive a membership for the entire cluster of 189\,$\pm$\,29 objects, of which 22\,$\pm$\,19 are beyond the HBL.\\
Regarding the crude estimate of the substellar population we note that studies of clusters at distances around and beyond 1\,kpc will be hampered by an overwhelming background contamination at the faintest magnitudes (that cannot be reduced by deeper observations). Nevertheless, the estimated fraction of substellar objects in NGC\,2316 (2-22\,\%) is comparable, within the errors, to that derived for the Trapezium cluster (20-25\,\%\, Muench at al., 2002) and IC\,348 (14-20\,\%\, Muench et al., 2003) from a similar KLF analysis. This is additional evidence that brown dwarfs do not make a significant contribution to the overall cluster, either by numbers or mass. In other words, the population of these clusters is dominated by stellar rather than substellar members.\\
The NIR excess emission obtained for NGC\,2316 from $JHK$ photometry is likely  an underestimate. Significant emission of a circumstellar disk is in the $L$--band (3.4\,$\mu$m), arising from warm circumstellar material within a few stellar 
radii, so $JHKL$ photometry would provide a more complete census of excesses, 
(\cite{Liu}). IC\,348, a cluster of a\-ppro\-xi\-ma\-tely 3\,Myr, presents a 
fraction of NIR excess sources of 21\% using $JHK$ photometry and 65\% 
using $JHKL$ pho\-to\-me\-try, (\cite{haisch,lada3}).
On the other hand, NGC\,2316 has a bright UV source, a central B3 star, 
which might accelerate the dissipation of circumstellar disks by 
photoevaporation and consequently decrease the number of observed 
sources with 
NIR excesses.\\
The comparison of the NGC\,2316, IC\,348 and Trapezium KLFs in section 4.4 shows good agreement between their shapes; in particular, the first two are very similar (apart from the number of sources), as they roughly show the same broad peak at the same absolute magnitudes. Assuming an universal IMF, (e.g. \cite{Lada2}), we infer that NGC\,2316 has approximately the same age as IC\,348, about 2--3\,Myr (Lada \& Lada, 1995; Muench et al., 2003), but is more evolved and older than the Trapezium.

\begin{acknowledgements}
We thank Lynne Hillenbrand for helpful discussion.
This research is financially supported by Funda\c{c}\~ao para a 
Ci\^encia e Tecnologia (FCT), Portugal, under the project 
\emph{PESO/P/PRO/40154/2000}. J. C. Correia gratefully acknowledges the financial 
support from FCT through the grant \emph{SFRH/BPD/3614/2000}.

\end{acknowledgements}

\end{document}